\documentclass[letterpaper,twocolumn,10pt]{article}
\usepackage{usenix-2020-09}

\usepackage{tikz}
\usepackage{amsmath}
\usepackage{enumitem}
\usepackage{float}
\usepackage{subcaption}
\usepackage{graphicx}

\begin{document}

\date{}

\title{\Large \bf Unique Identification of 50,000+ Virtual Reality Users\\from Head \& Hand Motion Data}

\author{
\and
\and
{\rm Vivek Nair}\\
UC Berkeley
\and
{\rm Wenbo Guo}\\
UC Berkeley
\and
{\rm Justus Mattern}\\
RWTH Aachen
\and
{\rm Rui Wang}\\
UC Berkeley
\and
\and
\and
{\rm James F. O'Brien}\\
UC Berkeley
\and
{\rm Louis Rosenberg}\\
Unanimous AI
\and
{\rm Dawn Song}\\
UC Berkeley
} 

\maketitle

\begin{abstract}

With the recent explosive growth of interest and investment in virtual reality (VR) and the so-called ``metaverse,'' public attention has rightly shifted toward the unique security and privacy threats that these platforms may pose. While it has long been known that people reveal information about themselves via their motion, the extent to which this makes an individual globally identifiable within virtual reality has not yet been widely understood. In this study, we show that a large number of real VR users (N=55,541) can be uniquely and reliably identified across multiple sessions using just their head and hand motion relative to virtual objects. After training a classification model on 5 minutes of data per person, a user can be uniquely identified amongst the entire pool of 50,000+ with 94.33\% accuracy from 100 seconds of motion, and with 73.20\% accuracy from just 10 seconds of motion. This work is the first to truly demonstrate the extent to which biomechanics may serve as a unique identifier in VR, on par with widely used biometrics such as facial or fingerprint recognition.

\end{abstract}

\section{Introduction}
\label{sec:introduction}

The emergence of affordable standalone virtual reality (VR) devices, such as the Meta Quest 2, has allowed VR to reach mass-market adoption in recent years, with nearly 10 million VR headsets sold in 2022 alone \cite{vr_sales}.
Coinciding with this dramatic increase in VR usage is a wave of new academic research revealing a range of unique security and privacy threats associated with these devices \cite{garrido_sok_2023}.

Gaming has thus far been the predominant driver of VR adoption, with 91 of the 100 most popular VR applications being games as of early 2023 \cite{steam_most_used}.
While gaming is typically perceived as a fairly innocuous class of applications from a privacy standpoint, the opposite may actually be true in VR. In this paper, we examine the extent to which spatial telemetry captured during VR gaming sessions can be used to uniquely identify an otherwise anonymous player.

We obtained a novel dataset containing over 2.5 million recordings of users playing ``Beat Saber,'' a VR rhythm game that today is by far the most popular VR application \cite{wobbeking_beat_2022}. Using a unique combination of context-aware featurization and hierarchical machine learning, players can be identified out of a pool of over 50,000 candidates with 94.33\% accuracy from 100 seconds of head and hand motion data, or with 73.20\% accuracy from just 10 seconds of movement.

It has long been understood that individuals exhibit distinct biomechanical motion patterns that can be used to identify them or infer their personal attributes \cite{Obrien:2000:AJP, Kirk:2005:SPE, cutting_recognizing_1977, kozlowski_recognizing_1977, pollick_gender_2005, jain_is_2016}. However, the extent to which the subset of this information that is observable in VR can be used to uniquely identify users is less well understood.
Although prior research has been conducted on the personal identifiability of VR tracking data \cite{pfeuffer_behavioural_2019, pfeuffer_behavioural_2019, liebers_understanding_2021, tricomi_you_2022, nair_exploring_2022}, existing works have utilized data from small lab studies with $16$ to $511$ participants. By contrast, our dataset is not only more than 100 times larger than the largest prior result, but is also far more representative of a realistic use case, comprising 55,541 real VR users across over 40 countries and using over 20 different types of VR devices.

Despite the difficulty of identification growing in proportion to the number of users, we achieve comparable identification accuracy to the prior works. We show that while identifying users in smaller sets ($\leq$~511) can be accomplished just by learning static attributes like height, actual behavioral differences in movement patterns must be utilized to identify users within such a large dataset. 
As such, we argue that our work is the first to truly demonstrate the extent to which motion can be an identifying feature in VR. \\

\noindent \textbf{Contributions:}
\vspace{-0.25em}
\begin{enumerate}[leftmargin=*,itemsep=0em]
    \item We have identified the largest and most representative dataset to date of virtual reality telemetry recordings (\S\ref{sec:dataset}).
    \item Our featurization technique uses VR application context information to enhance VR user identification (\S\ref{sec:featurization}).
    \item Our hierarchical classification approach allows us to build a scalable identification model with 50,000+ classes (\S\ref{sec:identification}).
    \item We achieve 94.33\% identification accuracy across 55,541 users (\S\ref{sec:evaluation}) and provide detailed explainability results (\S\ref{sec:explanations}).
\end{enumerate}
\clearpage

\section{Background}
\label{sec:background}

A virtual reality device uses an array of sensors to generate a stream of information about its user, which is consumed by an onboard or external computer to render stimuli for the user, thereby creating an immersive experience. In the case of multi-player (or ``metaverse'') applications, the generated data is also shared with a variety of external systems, which could then use it to infer private user information.

The 2023 VR privacy SoK by Garrido et al. \cite{garrido_sok_2023} presents a standard model of VR information flow and threat actors, which we will briefly recount below and use to position our study within the broader landscape of VR privacy research.

\subsection{VR Information Flow}

A typical consumer-grade virtual reality system comprises at least a head-mounted display (HMD) and two hand-held controllers. The system uses either external or onboard sensors to measure the position and orientation of these devices in 3D space, providing six degrees of freedom (6DoF) per tracked object. These six measurements per object are taken for the user's head and hands, constituting 18 tracked dimensions in total. The data are captured at a usual rate of between 60 and 144 times per second, resulting in a ``telemetry stream.''

Many VR devices contain additional sensors, such as microphones, cameras, and eye or full-body tracking devices. In this paper, we focus entirely on the basic motion telemetry data noted above, so as to investigate the question of how users can be identified by their motion alone.

The telemetry stream generated by a VR device is first used by a client-side application running on an onboard or connected computer to render a separate series of visual stimuli (or ``frames'') for each eye, along with auditory and haptic stimuli, creating an immersive 3D virtual world.

In the case of a metaverse experience, the application also forwards the telemetry stream to an external game server. The server, in turn, forwards this telemetry to other connected users, so that a virtual representation (or ``avatar'') of each user can be rendered on the devices of all other users.

\subsection{VR Threat Model}
Per the Garrido VR threat model, each entity in the above information flow that can view the VR device telemetry of a target user is considered a potential adversary. Specifically, the attackers generally considered in VR privacy research are VR hardware (I), VR applications (II), external servers (III), and external users (IV). Each of these adversaries receives a view of the telemetry stream, which it could use to make adversarial inferences of private VR user information instead of (or in addition to) its intended purpose of facilitating application functionality. However, because the data can be reduced and compressed at each stage of the information flow, adversaries in higher tiers are considered ``weaker'' in this model.

\begin{figure}[h]
\includegraphics[width=0.75 \linewidth]{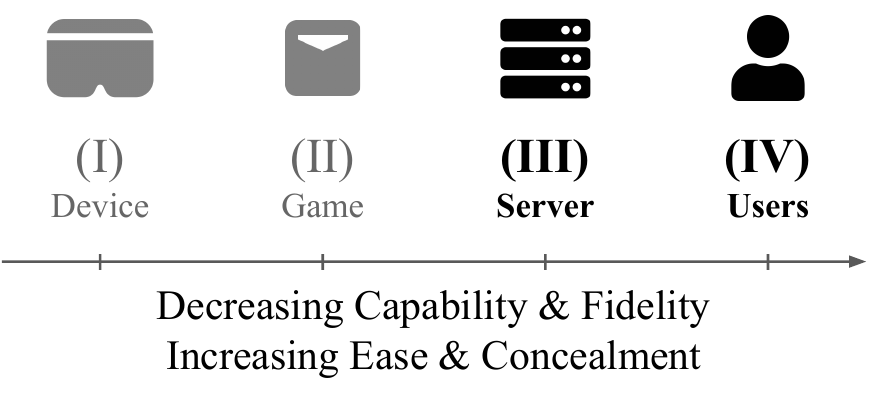}
\centering
\caption{Selected VR threats relevant to this work.}
\label{fig:threat_model}
\end{figure}

Fig. \ref{fig:threat_model} illustrates the general information flow and threat actors discussed thus far. In this paper, we are particularly interested in the game server (III) and other users (IV) as potential adversaries. These parties receive data processed by and filtered through the prior entities, meaning that attacks available to them can often be performed by other entities with even greater precision. They are also amongst the hardest attacks to detect due to their remote nature. This study exclusively analyzes data sent from a popular VR game to a remote server or other users, meaning that our attacks represent the hardest and most pernicious realistic threats in VR.

\subsection{Beat Saber}
``Beat Saber'' \cite{beat_saber} is an award-winning virtual reality rhythm game where players slice blocks representing musical beats with a pair of sabers they hold in each hand. With over \mbox{6.2 million} copies sold, Beat Saber is the most popular and highest-grossing VR game of all time \cite{wobbeking_beat_2022}.

\begin{figure}[H]
\includegraphics[width=\linewidth]{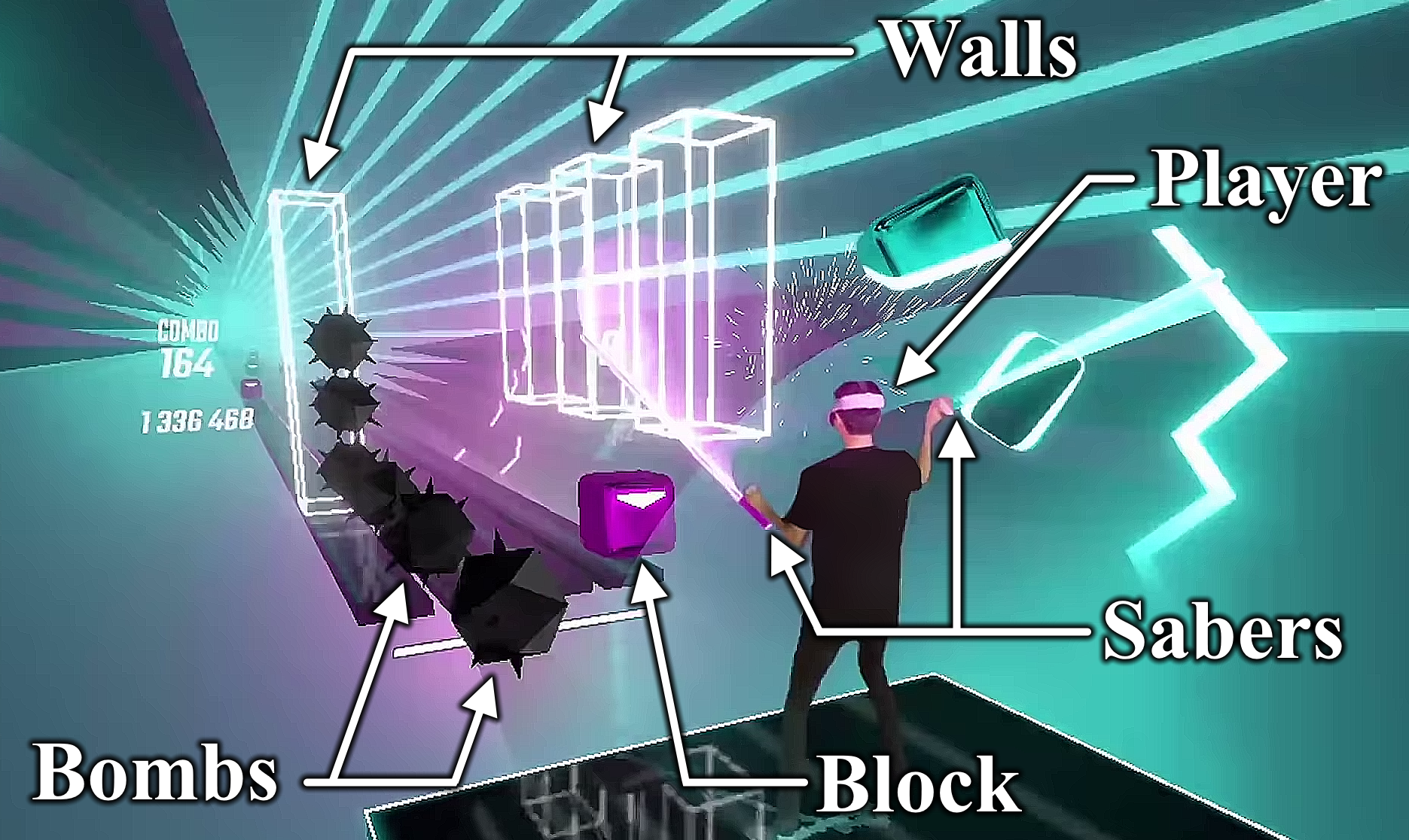}
\centering
\caption{Interactive objects in Beat Saber game.}
\label{fig:beat_saber}
\end{figure}

The Beat Saber game is segmented into ``maps'' which consist of a music track and a series of challenges presented to the user in time with the music. These challenges include ``blocks,'' which the player must hit with the correct saber (as indicated by the color) and angle (as indicated by the arrow), ``bombs,'' which the player must avoid hitting with their sabers, and ``walls,'' which the player must avoid with their head (see Fig.~\ref{fig:beat_saber}). If the player completes these actions successfully, they are awarded points according to their level of accuracy. On the other hand, if the player performs poorly, the map may terminate early with a ``level failed'' message.

While users can play and compete for high scores on hundreds of maps included in the base game or as purchasable extensions, over 100,000 user-created maps for popular songs are available by installing open-source game modifications.
\vspace{-0.5em}

\section{Dataset}
\label{sec:dataset}
``BeatLeader'' \cite{beatleader} is an open-source Beat Saber extension that maintains third-party leaderboards for over 100,000 custom Beat Saber maps. Beat Saber players may choose to install the BeatLeader extension in order to compete with other players to achieve a higher ``rank'' on the BeatLeader leaderboards. After playing a Beat Saber map with the BeatLeader extension enabled, scores are automatically uploaded to a globally-visible leaderboard. Since May 2022, over 50,000 users have posted over 2.5 million scores to the BeatLeader platform.

When uploading a score to BeatLeader, a recording (or ``replay'') of the user's performance is automatically captured by the BeatLeader extension and attached to their submission. The replay is made available to other BeatLeader users, who can use it to verify the authenticity of the submitted score.

In partnership with the administrators of BeatLeader, we obtained a 3.96~TB dataset consisting of 2,669,886 replays from 55,541 users across 713,013 separate play sessions. The dataset has between 1 and 4,509 replays per user, with a median of 14. The replays range in length from 5 seconds to over an hour \footnote{Some maps are longer than a single song; e.g., an entire film soundtrack.}, with a median length of 2 minutes and 56 seconds.

\subsection{Replay Format}
Replays in the dataset are encoded in the ``Beat Saber Open Replay'' (BSOR) \cite{bsor} format. BSOR files comprise four parts:

\vspace{-0.4em}

\begin{enumerate}[leftmargin=*,itemsep=-0.1em]
    \item \textbf{Metadata.} Device information and the values of all user-configurable game settings are included in the replay.
    \item \textbf{Telemetry.} The position and orientation of the player's head and hands is recorded every time a frame is rendered by the game, usually 60 to 144 times per second.
    \item \textbf{Context.} Replays encode the type, location, and timing of in-game stimuli, such as ``blocks,'' ``walls,'' and ``bombs,'' which the player is responding to throughout the replay.
    \item \textbf{Performance.} BSOR files also measure the validity and accuracy of the user's response to each in-game stimulus. 
\end{enumerate}

\subsection{User Attributes}
\label{sec:attr}

As noted above, Beat Saber replay files include metadata that reveals a number of user-specific data attributes. These attributes primarily consist of device information, such as the VR platform, runtime environment, software version numbers, and the make and model of the VR headset and controllers. They also contain the settings chosen by the user, including self-selected height and handedness. Finally, the number of replays present for each user indicates their level of ``experience,'' while their performance relative to other users playing the same map provides a measure of ``skill.'' The distribution of the users across these attributes is included in \S\ref{app:distribution}.

Because the goal of this paper is to uniquely identify VR users across sessions based on their motion alone, we do not incorporate any of the provided metadata into our identification models. However, we do use this information in \S\ref{sec:impact} to identify which attributes correlate with higher or lower identification accuracy and contextualize the results accordingly.


\vspace{-0.5em}

\subsection{Ethical Considerations}
Because our work involves data derived from human subjects, significant attention was given to ethics throughout the study. We note that no original data collection was performed by the authors; we used an existing dataset from an external source.

BeatLeader users must go out of their way to voluntarily modify their Beat Saber installation to add the BeatLeader extension and share their Beat Saber replay data. They are fully aware of the nature of the data being shared, as uploading and sharing Beat Saber replay data is the explicit purpose of the extension. They also consent to their replay data being used for a variety of purposes, including for data analysis, in the BeatLeader Privacy Policy. Prior to this study, the replay data was already publicly viewable to other BeatLeader users.

In the interest of user privacy, we have not publicly released the entire raw dataset used to produce this work, despite its already public nature. However, we will release enough de-identified and normalized data to replicate our results to any qualified researcher upon reasonable request. We also made no attempt in this work to infer user data that could be viewed as particularly sensitive, such as health information.

Overall, we believe this research constitutes a net benefit to society by highlighting the magnitude of the VR privacy threat and motivating future work on defensive countermeasures. It further benefits the Beat Saber users whose data was utilized by enabling the future development of powerful anti-cheating tools.
Since no original data was collected from human subjects, and BeatLeader data is already public, our study has been deemed IRB-exempt under 45 C.F.R. § 46.104(d)(4)(i).
\section{Related Work}
\label{sec:related}
\vspace{-0.5em}

\subsection{Motion}

Since as early as the 1970s, researchers have demonstrated that people reveal identifying information about themselves via their motion. In a 1977 study of 6 participants, Cutting and Kozlowski demonstrated that individuals can identify their friends with 38\% accuracy by viewing the motion of 8 tracked objects affixed to the body \cite{cutting_recognizing_1977}, and that the gender of the 6 participants could be identified by a stranger with 63\% accuracy using the same 8 tracked objects \cite{kozlowski_recognizing_1977}.

More recently, Pollick et al. (2005) \cite{pollick_gender_2005} have used statistical techniques to achieve 79\% accurate identification of gender from motion. In a study of 8 participants, Jain et al. (2016) \cite{jain_is_2016} found that the motion of children can be differentiated from that of adults with 66\% accuracy.

In two further related works, O'Brien et al. (2000) \cite{Obrien:2000:AJP} and Kirk et al. (2005) \cite{Kirk:2005:SPE} demonstrated the ability to use motion data to infer a person's skeletal structure. O'Brien et al. used 16 sensors recorded with 6 degrees of freedom, while Kirk et al. used 30 to 40 optical markers captured with 3 degrees of freedom. Although not the explicit purpose of these works, the skeletal models could be used for user identification.

Virtual reality is somewhat distinct from the situations described above in that only 3 tracked locations are typically provided rather than the 8 to 40 used in the mentioned studies.
Until the relatively recent proliferation of VR technology, the applicability of these results to VR has been uncertain.

\vspace{-0.5em}

\subsection{Virtual Reality}

The 2023 Garrido et al. VR privacy SoK \cite{garrido_sok_2023} provides a recent survey of the VR privacy research landscape. Our work would fall under the ``geospatial telemetry'' attribute class in the SoK's taxonomy. Here, we summarize the works listed in the same category which are most relevant to our own.

First, Pfeuffer et al. (2019) \cite{pfeuffer_behavioural_2019} performed a laboratory study of 22 users, who were instructed to perform a variety of tasks in VR (pointing, grabbing, walking, typing) across two sessions. Using a random forest model, they were able to identify a user within the set of 22 with up to 40\% accuracy.

Next, Miller et al. (2020) \cite{pfeuffer_behavioural_2019} conducted a lab study of 511 users, whose telemetry was captured while they watched a series of 360-degree videos in VR. With a random forest model, their system correctly identifies users within the pool of 511 with 95\% accuracy from 5 minutes of telemetry data.

Liebers et al. (2021) \cite{liebers_understanding_2021} conducted a similar lab study of 16 users, who were asked to play archery and bowling games in VR. They were able to identify users within the set of 16 using an LSTM model with 90\% accuracy.

Finally, Tricomi et al. (2022) \cite{tricomi_you_2022} demonstrated the profiling of AR and VR users with laboratory studies of 34 and 35 users, respectively. They uniquely identify 30 users in VR with 95\% accuracy using a logistic regression model.

Overall, Miller et al. is the largest known study of VR user identifiability, with 511 users across 5,110 sessions. Our study, with 55,541 users and 713,013 sessions, is thus at least two orders of magnitude larger than the largest existing result.

Furthermore, while all of the above works involve data collected from a highly-controlled laboratory setting with 1 to 3 device types, our dataset originates from real VR users in 40+ countries, and includes 20+ types of VR devices.

Despite the significantly harder task of identifying users amongst tens of thousands of possible classifications and in uncontrolled environments, we achieve comparable accuracy to the prior works. We believe this is the first study to truly demonstrate the staggering scale of the VR privacy threat.

\vspace{-0.5em}

\subsection{Machine Learning}
\label{sec:ml_rw}

\noindent{\textbf{Classical ML.}}
As summarized above, existing VR privacy studies model user identification as a classification problem and leverage machine learning to classify users based on feature vectors of extracted data.
Given that the existing studies process the telemetry data into a relatively small tabular dataset, these works usually leverage classical ML techniques (such as random forest~\cite{breiman2001random} and gradient boosting~\cite{chen2015xgboost}).

Underlying these models are decision trees, which construct a tree-based rule structure for a learning problem.
A random forest ensembles multiple decision trees to improve the model's capacity, and thus is capable of handling more sophisticated learning problems.
Gradient boosting takes this a step further by iteratively optimizing the set of trees rather than simply aggregating them.
During the training process, gradient boosting actively updates the trees and their weights based on the current prediction results, allowing it to generally achieve a better performance than random forests alone \cite{clark2017tree}.
We observe similar results in our study, with gradient boosting models providing by far the best performance.

\smallskip

\noindent{\textbf{Deep Learning.}}
Interestingly, only one of the existing studies (Liebers et al. \cite{liebers_understanding_2021}) has used deep learning-related techniques for user identification, and its results are amongst the least accurate at 90\% accuracy with 16 users.
This is counterintuitive, as deep learning has become a mainstream technique in the machine learning community. 
Research in different application domains has demonstrated that deep learning algorithms (e.g., Multi-layer Perceptrons), outperform traditional (e.g., tree-based) ML models in dealing with tabular data~\cite{goodfellow2016deep}.

However, this may not be the case in VR user identification. 
This application has a very large number of users, which means that the classifier has to distinguish a large number of classes.
It is challenging for deep learning models to train and converge under these conditions because they require a multi-class classifier to contain a large number of neurons in the output layer. 
In fact, most existing benchmark datasets where deep learning demonstrates a superior performance have a small number of classes.
For example, the widely used image classification datasets MNIST~\cite{deng2012mnist} and CIFAR-10~\cite{krizhevsky2010convolutional} have ten classes, and some widely used text classification datasets only have 20 classes (Newsgroups~\cite{albishre2015effective}).
The dataset with the most classes is ImageNet~\cite{deng2009imagenet}, which has 1,000 classes.

We found that deep learning empirically fails to perform well in our study, which requires more than 50,000 classes. Still, it is likely that larger and more sophisticated deep learning models could achieve strong performance in the future.
\section{Featurization}
\label{sec:featurization}

In this section, we describe our method for converting the time-series replay telemetry data into a flat feature vector which can be consumed by a basic non-sequential model. The featurization techniques described in this section are used in the identification models discussed later in this paper.

We define a ``session'' as a continuously-recorded sequence of replays from a single user where no more than 10 minutes have elapsed between each replay. Our dataset contains an average of 13 such sessions per user. For each user, we reserve 70\% of the sessions for training, 10\% for validation, and 20\% for testing, with a minimum of 1 session per set. As such, our models always perform true cross-session user identification rather than merely learning session-specific features, such as the exact position of a user within their room.

We begin with the best-performing existing method of featurizing VR telemetry data,  which is that of Miller et al. \cite{miller_personal_2020}, achieving 95\% accuracy on 511 users. We describe this method in \S\ref{sec:motion}, and improve upon it in subsequent parts.

Throughout this section, we use a 500-user identification model to validate our featurization choices and compare the resulting classification accuracy to the Miller et al. approach. For each proposed featurization approach, we randomly chose 500 users from our dataset and generated 150 training and 15 testing samples per user, using the train/test split discussed above. The features were then standardized using Z-score normalization before being used to train a 500-class LightGBM classification model. The identification accuracy on a per-sample and per-user basis is used to evaluate each approach.

\subsection{Guiding Principles}

\begin{figure}[H]
\begin{subfigure}[h]{0.33\linewidth}
\includegraphics[width=\linewidth]{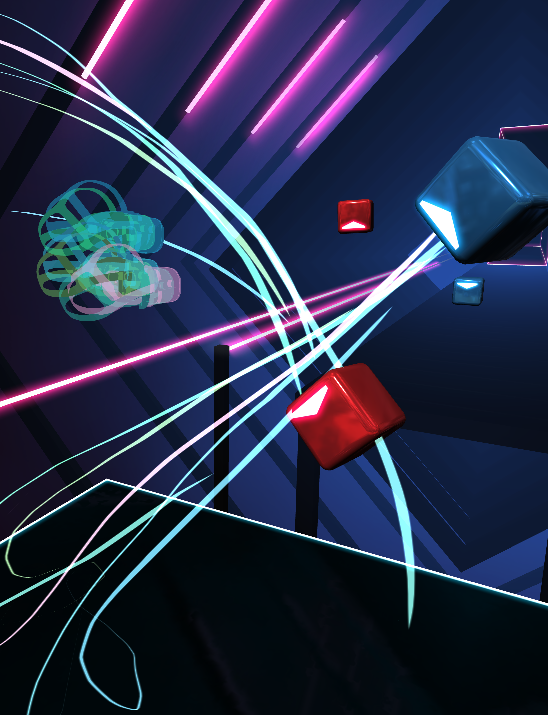}
\end{subfigure}%
\hfill
\begin{subfigure}[h]{0.33\linewidth}
\includegraphics[width=\linewidth]{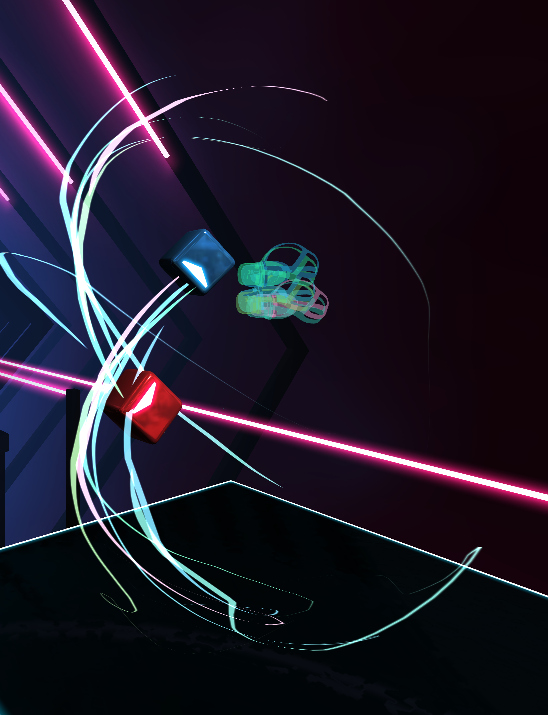}
\end{subfigure}%
\hfill
\begin{subfigure}[h]{0.33\linewidth}
\includegraphics[width=\linewidth]{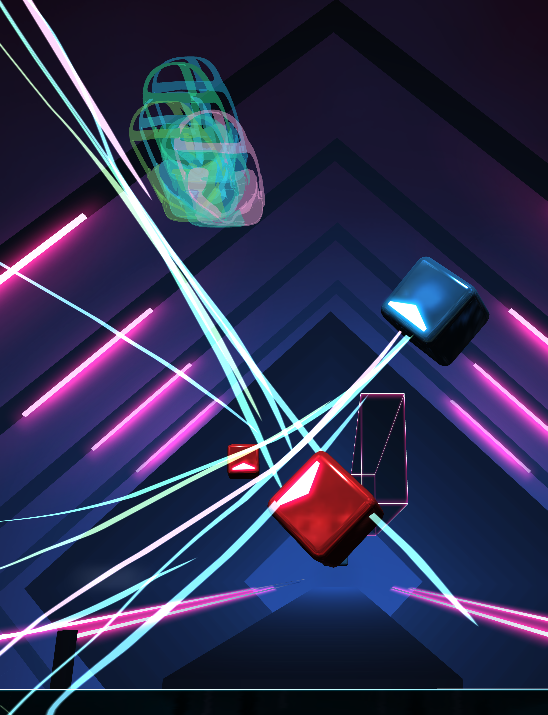}
\end{subfigure}%
\caption{Five Beat Saber users hitting the same block pattern.}
\label{fig:comparison}
\end{figure}

Fig. \ref{fig:comparison} shows, from several perspectives, the path taken by five Beat Saber users when slicing the same pair of blocks. As is clearly visible by the depictions, different users exhibit distinct motion responses even when presented with identical stimuli.
These differences may be the result of physiology, learned motion patterns (``muscle memory''), random variance, or a combination thereof.
The goal of the identification models presented in this paper is to learn a set of motion characteristics that uniquely represent a user. Accordingly, the featurization techniques of this section aim to reduce the dimensionality of the telemetry stream to the extent possible while retaining the ability to differentiate between users.

\subsection{Motion Features}
\label{sec:motion}

\begin{figure}[H]
\includegraphics[width=\linewidth]{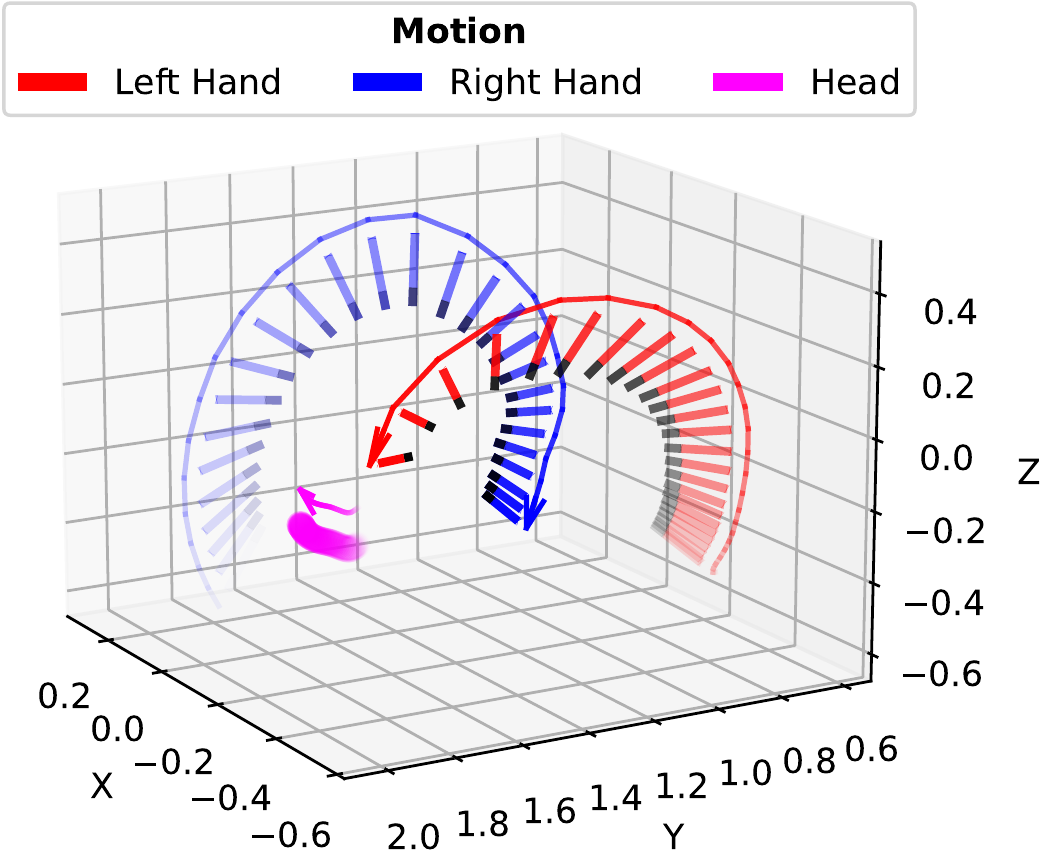}
\centering
\caption{Head and hand motion from one second of telemetry.}
\label{fig:motion}
\end{figure}

Motion data (telemetry) is the primary source of data for user identification and inference in VR. Fig. \ref{fig:motion} shows a one-second segment of the head and hand motion of a Beat Saber user. 

As is visible in Fig. \ref{fig:motion}, each frame of telemetry data encodes 3D position and orientation coordinates across each of the three tracked objects. The Miller et al. method of motion data encoding suggests summarizing each of these 18 data streams using five summary statistics, namely the minimum, maximum, mean, median, and standard deviation, resulting in a 90-dimensional output vector. Using this approach with the Beat Saber data yields a 69.3\% accurate per-sample identification and 93.4\% accurate per-user identification using the evaluation method described above. This is  comparable to the 95\% accuracy reported by Miller et al. with their dataset.

In practice, we found that better performance is achieved by providing orientation measurements as four quaternion elements instead of three Euler angles. This modification alone resulted in an improved per-sample identification accuracy of 80.1\% and per-user identification accuracy of 96.6\%. Thus, our best-performing motion featurization can be represented as a 105-dimensional vector constructed as follows:

\begin{center}
$\{\mathit{pos}_x,\mathit{pos}_y,\mathit{pos}_z,\mathit{rot}_i,\mathit{rot}_j,\mathit{rot}_k,\mathit{rot}_{1}\}$ \\
$\times$ \\
$\{\mathit{min},\mathit{max},\mathit{mean},\mathit{med},\mathit{stdev}\}$ \\
$\times$ \\
$\{\mathit{head},\mathit{left\_hand},\mathit{right\_hand}\}$
\end{center}

\subsection{Context Features}
\label{sec:context}

While motion alone may be sufficient to identify 500 users, additional information is needed when dealing with significantly larger datasets. In particular, models can benefit from knowing the activity-specific context in which a motion segment is captured such that different users can be compared directly when performing the same action.

\begin{figure}[H]
\includegraphics[width=\linewidth]{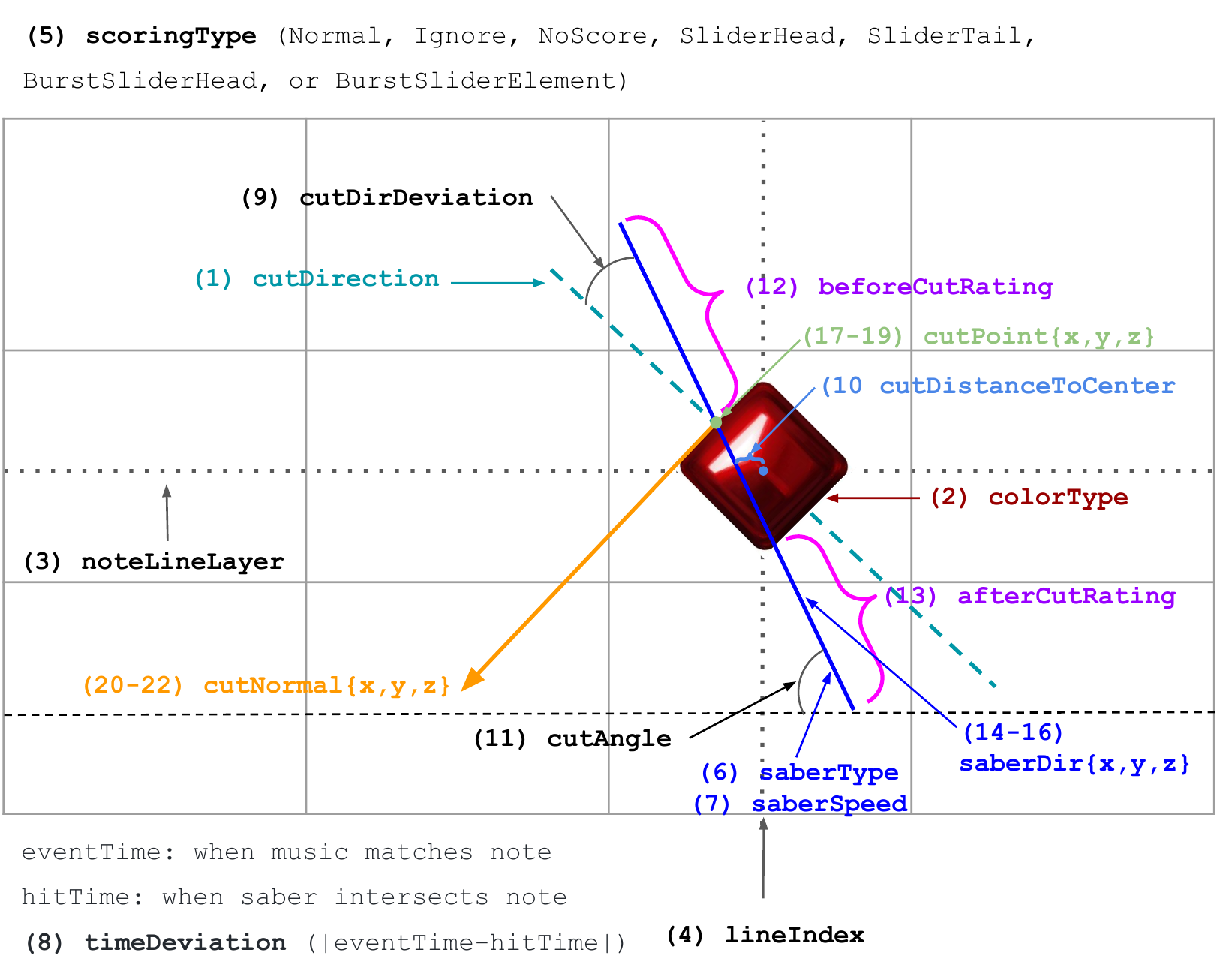}
\centering
\caption{The 22 contextual features of a Beat Saber block.}
\label{fig:context}
\end{figure}

In the case of Beat Saber, the activity chosen was the act of slicing an approaching block with a saber held in either hand. Specifically, we found 22 features that most accurately characterize movement relative to a single block, as shown in Fig. \ref{fig:context}. These features include, for example, the position, orientation, type, and color of the block, the angle, speed, location, and accuracy of the cut, and the relative error of the cut in both space and time.

Although these 22 features provide a comprehensive yet succinct parameterization of a user's response to an individual block, they are insufficient to identify users without accompanying motion features. Using these features alone with the previously-established evaluation method yields just 14.8\% accuracy per sample and 43.8\% accuracy per user.
While this is still highly statistically significant relative to the 0.2\% accuracy one would achieve by attempting to identify one of the 500 users at random, it under-performs even the basic Miller et al. approach. Still, it demonstrates the potential to aid identification when combined with motion features.

\subsection{Hybrid Featurization}
\label{sec:hybrid}

Finally, we describe the inclusion of both motion and context features within a single feature vector, thus allowing models to interpret motion data specifically in relation to other users performing the same or similar actions.
By combining the 22 context features of \S\ref{sec:context} with the 105 motion features of \S\ref{sec:motion} corresponding to one second of motion centered on the moment of contact, a 127-dimensional hybrid feature vector can be produced. Using this feature set with our established evaluation approach yields 83.8\% accurate per-note user identification, with 98.2\% accurate identification per user.

While this hybrid feature set now outperforms either the motion or the contextual features alone, some useful information is still excluded.
In particular, it is useful to explicitly separate the motion features from before and after a target event. For example, different information can be learned from a user's ``in swing'' and ``out swing'' relative to a block.

\begin{figure}[H]
\includegraphics[width=0.75 \linewidth]{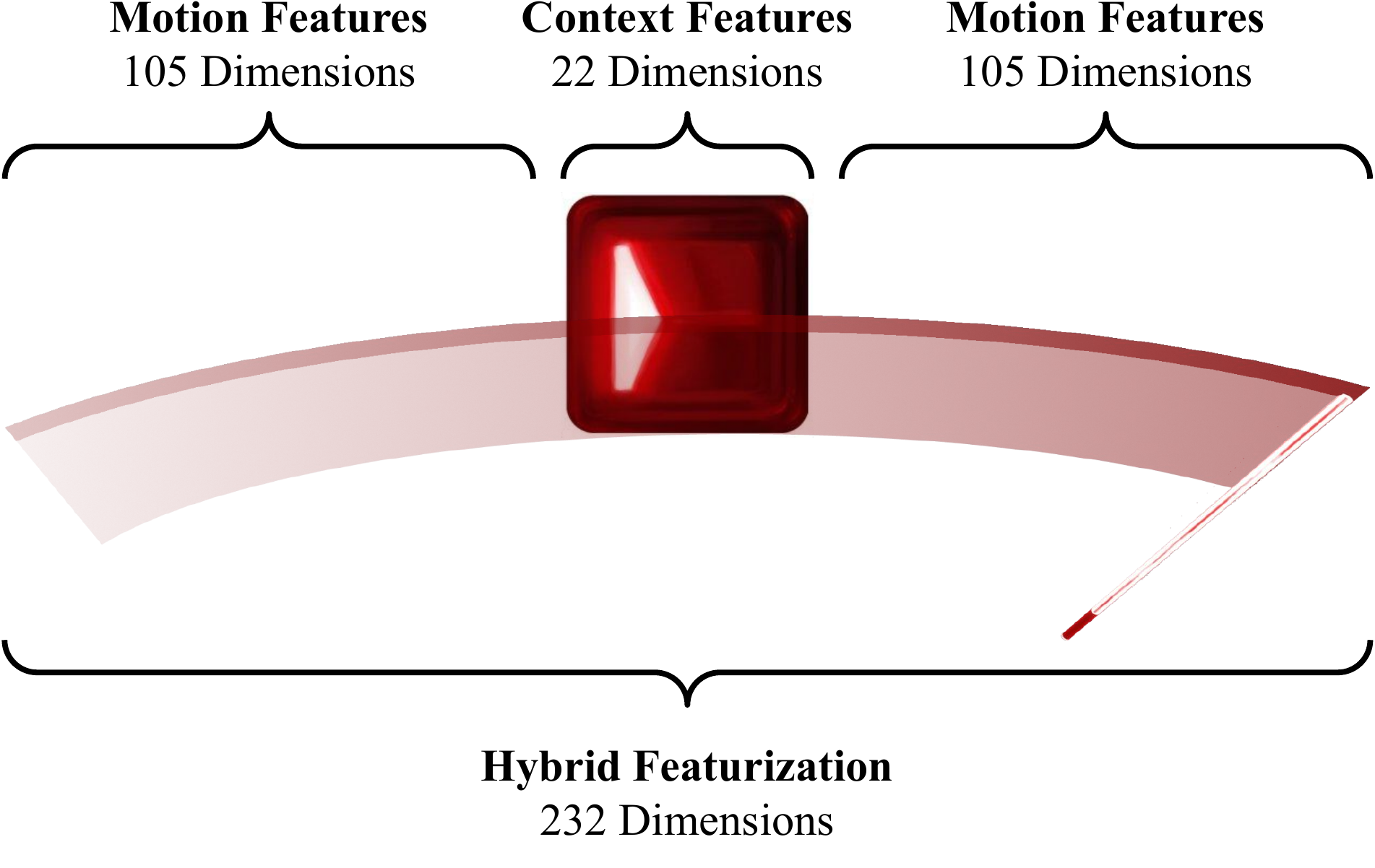}
\centering
\caption{Hybrid featurization of a Beat Saber block.}
\label{fig:hybrid}
\end{figure}

Fig. \ref{fig:hybrid} shows a full hybrid featurization of a Beat Saber block, including 22 contextual features for the block and 105 motion features corresponding to the one-second intervals before and after the block, totalling 232 dimensions. When evaluating this featurization with the same machine learning approach as before, 93.2\% accurate identification is achieved per sample, with perfect (100.0\% accurate) per-user identification of 500 users. The results of all approaches discussed in this section are summarized in Table \ref{tab:featurization}.

\begin{table}[H]
\resizebox{\columnwidth}{!}{%
\begin{tabular}{|c|c|c|c|}
\hline
\textbf{\begin{tabular}[c]{@{}c@{}}Featurization\\ Approach\end{tabular}} & \textbf{\begin{tabular}[c]{@{}c@{}}Features\\ (\#)\end{tabular}} & \textbf{\begin{tabular}[c]{@{}c@{}}Accuracy\\ (Per Sample)\end{tabular}} & \textbf{\begin{tabular}[c]{@{}c@{}}Accuracy\\ (Per User)\end{tabular}} \\ \hline
Motion (Euler Angles) & 90 & 69.3\% & 93.4\% \\ \hline
Motion (Quaternion) & 105 & 80.1\% & 96.6\% \\ \hline
Contextual & 22 & 14.8\% & 43.8\% \\ \hline
Light Hybrid & 127 & 83.8\% & 98.2\% \\ \hline
Full Hybrid & 232 & 93.2\% & 100.0\% \\ \hline
\end{tabular}%
}
\caption{Accuracy of identifying 500 users using LightGBM with each of the discussed featurization methods.}
\label{tab:featurization}
\end{table}

In summary, the combination of rich contextual information about an event with separate features summarizing motion before and after said event is effective at achieving accurate identification for datasets significantly larger than 500 users.
This is in part because the motion segments can be understood in the context of the corresponding stimuli, and in part because it begins to simulate a small sequential model; that is, it allows the model to ascertain which motion features are consistent and which change across two consecutive time slices.
As such, we use this 232-dimension hybrid featurization method in all subsequent models for the remainder of this paper.
\section{Model Architecture}
\label{sec:identification}

Having established the above featurization technique, we next describe our selected machine learning model architecture for identifying users.  This remains a non-trivial problem in practice, as it requires a 50,000-class classification model, a use case that most existing machine learning algorithms are not designed to handle (see \S\ref{sec:ml_rw}). Therefore, after selecting an optimal algorithm and preprocessing method, we describe a hierarchical approach for constructing the overall classification model out of several smaller classifiers.

\subsection{Algorithm Selection}
\label{sec:alg}

Using the best-performing feature set from \S\ref{sec:featurization}, we tried to construct an identification model using 6 popular classical machine learning classification algorithms with the same sample of 500 users. For each algorithm, we began by using the default hyperparameters and then ran up to 25 rounds of tuning to obtain the below results, which show the best per-sample identification performance achieved by each algorithm.

\begin{itemize}[leftmargin=*,itemsep=-0.2em]
    \item LightGBM: \textbf{93.2\%}
    \item XGBoost: 80.0\%
    \item Logistic Regression: 72.2\%
    \item Support Vector Machines: 67.13\%
    \item Extreme Random Trees: 35.5\%
    \item Random Forest: 32.1\%
    \item Naive Bayes: \textbf{1.2\%}
\end{itemize}

As discussed in \S\ref{sec:ml_rw}, gradient boosting models are known to outperform other tree-based classification algorithms on tabular datasets, which matches our observations above. In particular, LightGBM \cite{ke_lightgbm_2017}, an industry-leading gradient boosting framework, exhibited by far the best performance.

We also tried multiple sequential and non-sequential deep learning approaches with limited success. As summarized below, the deep learning attempts far underperformed the classification accuracy of the best classical ML algorithm.

\begin{itemize}[leftmargin=*,itemsep=-0.2em]
    \item GRU: \textbf{84.0\%}
    \item LSTM: 83.0\%
    \item MLP: \textbf{72.0\%}
\end{itemize}

Overall, we conclude that simple deep learning algorithms empirically failed to perform as well as LightGBM for the large multi-class classification task at hand. Moving forward, we use LightGBM for our identification models in view of the performance results and the fundamental factors favoring gradient boosting for this type of application.

\subsection{Preprocessing Method}

Using the hybrid featurization and LightGBM model with optimal hyperparameters (see \S\ref{app:params}), we evaluated five potential preprocessing methods, the results of which are shown below.

\begin{itemize}[leftmargin=*,itemsep=-0.2em]
    \item StandardScaler: \textbf{93.2\%}
    \item MinMaxScaler: 89.8\%
    \item MaxAbsScaler: 86.4\%
    \item SparseNormalizer: 83.5\%
    \item TruncatedSVD: \textbf{66.5\%}
\end{itemize}

The preprocessing approach with best results is standard scaling (Z-score normalization), whereby each feature is transformed by removing the mean and scaling to unit variance.

\subsection{Hierarchical Approach}
\label{sec:approach}

\begin{figure}[H]
\includegraphics[width=\linewidth]{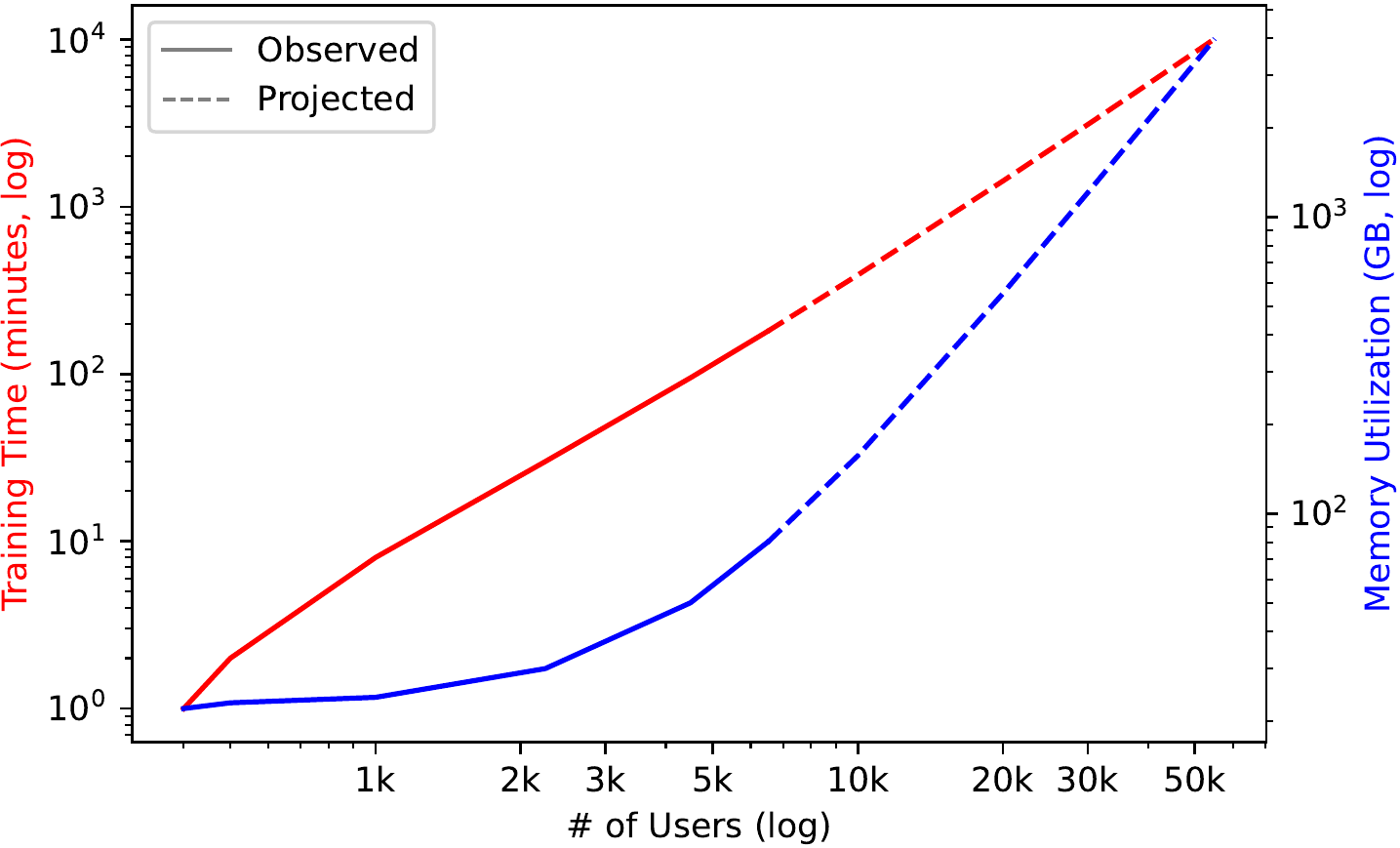}
\centering
\caption{Observed and projected time and memory required to train an increasingly large LightGBM classifier.}
\label{fig:scale}
\end{figure}

For smaller datasets, the above methods would be adequate. Indeed, if up to 5,500 classes were present, a single LightGBM classification model, deployed with our described featurization and preprocessing method, would demonstrate strong performance in identifying users. Unfortunately, training a single LightGBM model with 50,000 classes is infeasible with our dataset. We found that the training time and memory consumption of training a LightGBM classifier scales quadratically with the number of classes, as shown in Fig. \ref{fig:scale}.

According to a polynomial projection of our attempts to train classifiers with as many as 5,000 users, training a single classifier with all 55,000 users would take over 7 days and consume nearly 4~TB of RAM. While still within the realm of possibility when using server-grade hardware, the prospect of even larger datasets over the horizon motivates us to find a more efficient and scalable architecture.

We ultimately chose to construct a multi-layer hierarchical classifier. Our overall identification model is composed of three layers of smaller classifiers, each of which are only trained on a small set of available classes.

\begin{figure}[H]
\includegraphics[width=\linewidth]{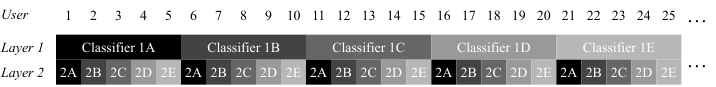}
\centering
\caption{Hierarchical structure with 5 models per layer.}
\label{fig:hierarchical_structure}
\end{figure}

Fig. \ref{fig:hierarchical_structure} illustrates the principle method by which the first two classification layers are constructed. In the first layer, N classifiers are each trained on 1/N of the available classes. In practice, we train 10 classification models with about 5,000 users each. This single layer already provides better performance than one may expect. Although each of the models will output a classification when identifying a user, regardless of whether that user is actually contained within their training set, the classification probability is usually highest in the model actually containing the target user.

\begin{figure}[H]
\includegraphics[width=\linewidth]{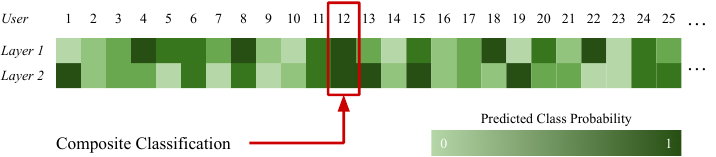}
\centering
\caption{Class probabilities output by hierarchical classifier.}
\label{fig:hierarchical_classification}
\end{figure}

Further accuracy can be obtained by adding a second layer, also containing N classifiers each trained on 1/N of the available classes, with an even class redistribution from the first layer. Now, when querying each layer to identify a user, the layers are likely to agree on the correct user while disagreeing about the false classification (see Fig. \ref{fig:hierarchical_classification}). The overall classification can now be obtained by taking the highest logarithmic sum of the class probabilities output by both layers.

\begin{figure}[H]
\includegraphics[width=\linewidth]{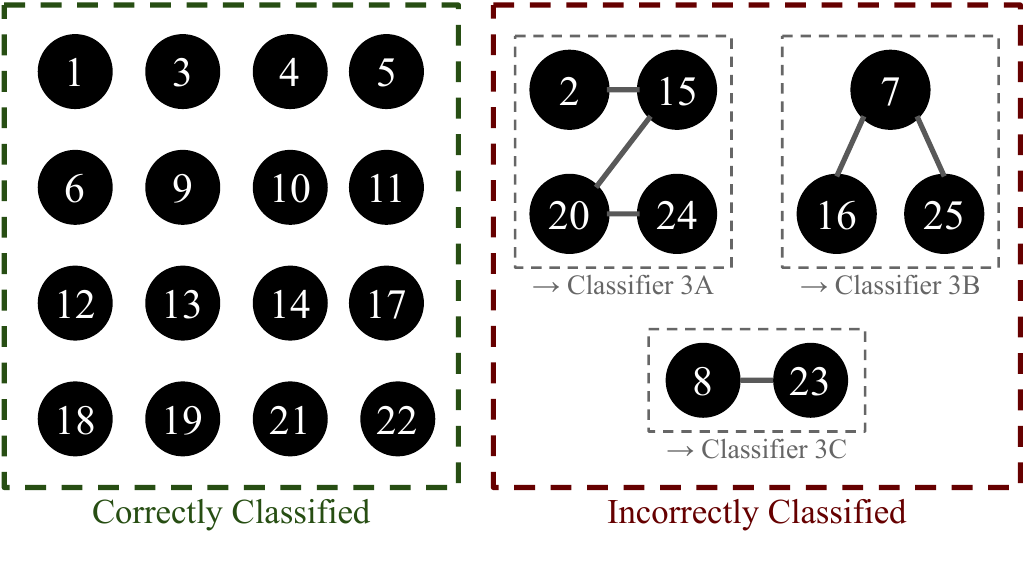}
\centering
\caption{Graph-based method of selecting layer 3 groups.}
\label{fig:hierarchical_components}
\end{figure}

Adding more layers at this stage via random redistribution provides diminishing returns. Instead, a separate clustering set (independent of the train, validate, and test sets) can now be used to cluster users based on their class confusion using the existing two layers. The method for doing so using connected components in a graph is illustrated in Fig. \ref{fig:hierarchical_components}.

As illustrated in Fig. \ref{fig:hierarchical_components}, an undirected graph is constructed with a node for each user. Every time a user is incorrectly classified using the clustering set, an edge is added between the user and up to five apparently similar users. The connected components of this graph now represent sets of users who are likely to be misidentified as each other. In a third layer, one additional model can be trained for each component $C$ in the graph (where $|C|>1$), containing the users of $C$.

When ultimately identifying a user, the logarithmic sum of the first two layers is used to obtain an initial identity. If the resulting user is present in one of the connected components, the corresponding model in the third layer is used to produce the final classification. Otherwise, the initial classification is directly returned as the predicted identity.
Given limited computational resources, this approach increases the odds that similar classes are directly compared in at least one model.

\subsection{Scalability}

While motivated initially by the infeasibility of training a single multiclass classification model of insufficient size, the proposed hierarchical architecture also presents a number of important scalability and practicality improvements over a monolithic approach. Each model in a layer can be trained in parallel, allowing for a 10-20x reduction in training time when using a cluster. Testing and inference can similarly be parallelized by evaluating each model separately.

Finally, the cost of adding a new user is significantly reduced by the hierarchical approach. When a new user is added, only one model on each layer must be retrained, rather than re-training the entire classifier. Given that most platforms where such an identification model may practically be deployed are constantly receiving new users, this alone constitutes a major improvement in the practicality of deployment.
\section{Evaluation}
\label{sec:evaluation}

We evaluated our identification technique using a distributed machine learning cluster of 10 nodes, each with 16~vCPU cores and 128~GB of RAM. The replays of each user were separated into 4 or more distinct sessions, which were reserved for training, clustering, validation, and testing at a ratio of 70-10-10-10. For each user, 150 samples were generated from the training set using the full hybrid featurization method of \S\ref{sec:hybrid}. The features of all users were then z-score normalized, and used to train the hierarchical model described in \S\ref{sec:approach}.

The training process was completed in about 3 hours each for the first and second layers and about 6 hours for the third layer. The final testing process, which required over 90 million classifications to be made, took about 8 hours; an individual user identification requires less than a second.

\subsection{Results}
\begin{table}[H]
\resizebox{\columnwidth}{!}{%
\begin{tabular}{|c|c|c|c|}
\hline
\textbf{Layer} & \textbf{\# of Models} & \textbf{\begin{tabular}[c]{@{}c@{}}Accuracy\\ (per Model)\end{tabular}} & \textbf{\begin{tabular}[c]{@{}c@{}}Accuracy\\ (per Layer)\end{tabular}} \\ \hline
Layer 1 & 10 & 93.1\% & 90.2\% \\ \hline
Layer 2 & 10 & 93.1\% & 90.2\% \\ \hline
Layers 1 \& 2 & 20 & 93.1\% & 91.0\% \\ \hline
Layer 3 & 5 & 84.0\% & 84.0\% \\ \hline
Layers 1, 2, \& 3 & 25 & 91.3\% & 94.3\% \\ \hline
\end{tabular}%
}
\caption{Accuracy of each hierarchical model layer per model (i.e., 5.5k users) and per layer (i.e., 55k users).}
\label{tab:results}
\end{table}

Table \ref{tab:results} shows the identification accuracy of each layer in the hierarchical model when evaluated using 50 test samples (100 seconds) per user. An identification accuracy of 90.1\% can be achieved using a single layer, with the hierarchical architecture boosting the overall accuracy to 94.3\%.

\begin{figure}[H]
\includegraphics[width=\linewidth]{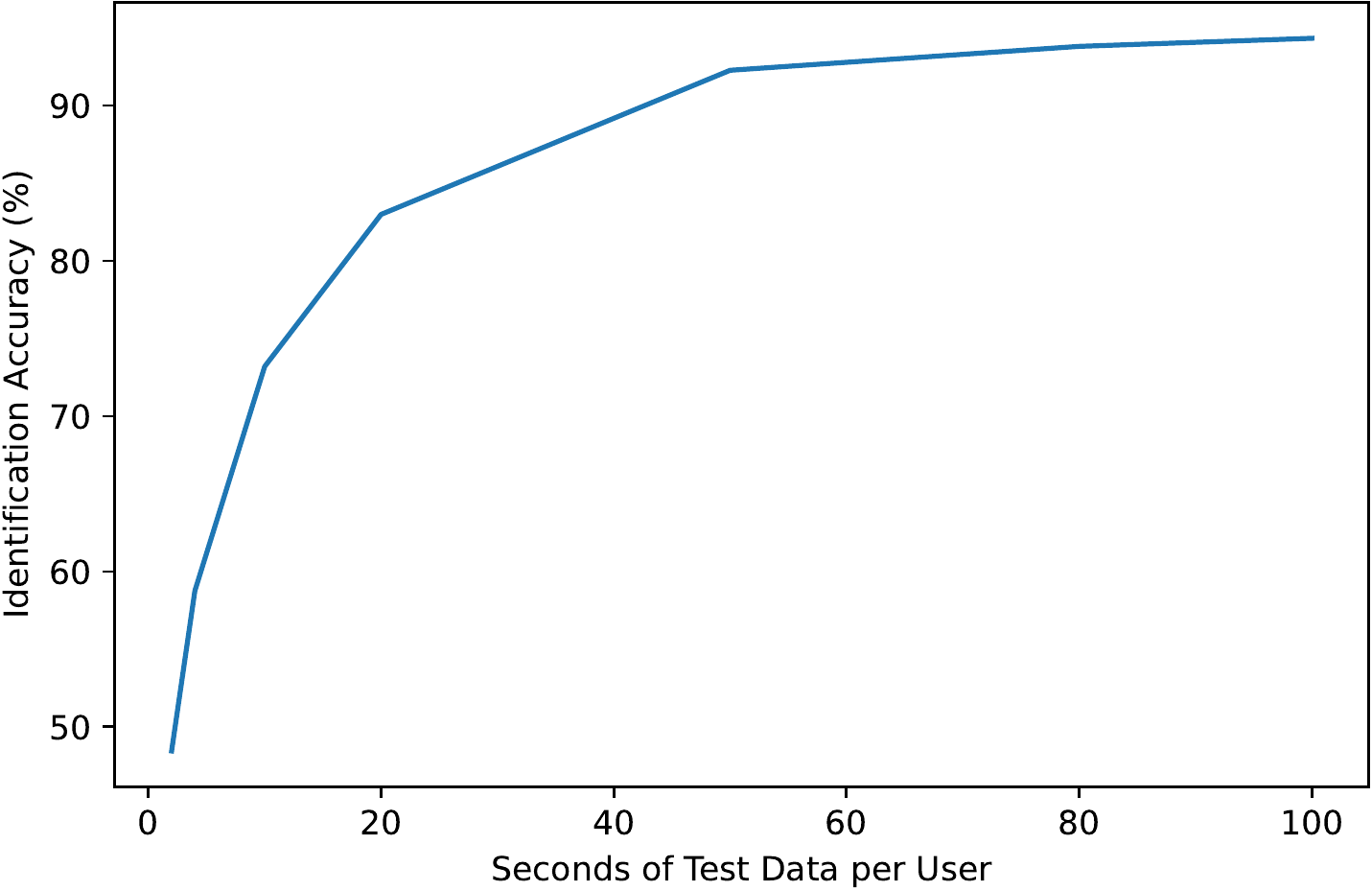}
\centering
\caption{Impact of test sample size on accuracy.}
\label{fig:samples}
\end{figure}

Of course, the accuracy of identification is highly dependent on the number of samples (and thus seconds of data) used to identify a user. Fig. \ref{fig:samples} illustrates the identification accuracy in relation to the number of seconds used. 

Even with a single sample generated from just 2 seconds of telemetry data, the correct user out of 50,000 is identified about 48.45\% of the time. Using 5 samples (10 seconds) of data increases this accuracy to 73.20\%, which indicates that only a short period of motion information is actually needed to uniquely characterize a user.
A single minute of data yields 92.78\% identification accuracy, and the full 94.33\% accuracy is achieved when 50 samples (100 seconds) of data are used, with rapidly diminishing returns for each sample thereafter.

In some cases, it may be sufficient to output a small number of candidate identities rather than exactly identifying a user. In our evaluation, the correct user is amongst the top 3 candidates identified by the model in 97.25\% of all instances.
    
\subsection{Impact Factors}
\label{sec:impact}

\begin{figure}[H]
\includegraphics[width=\linewidth]{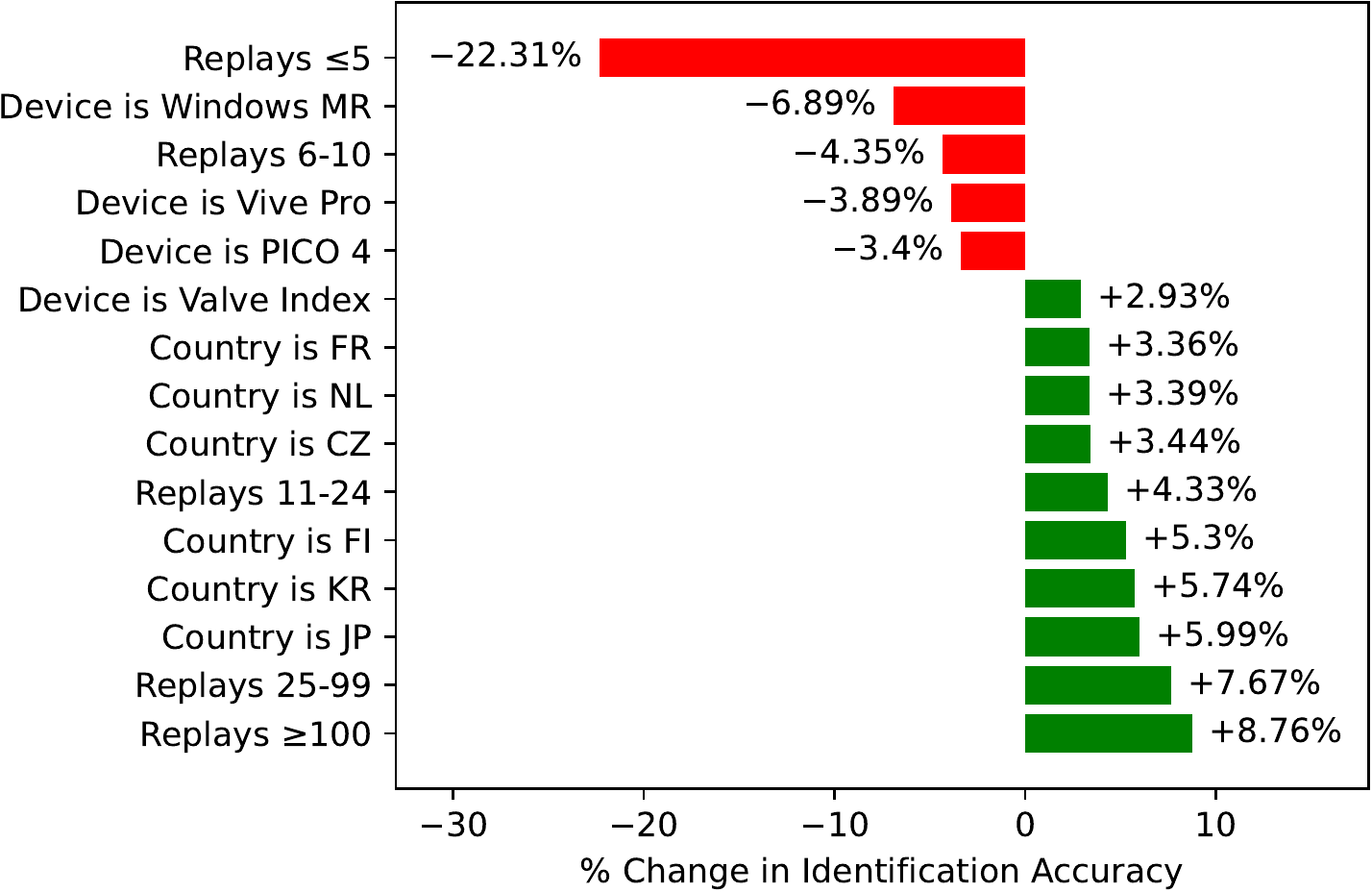}
\centering
\caption{Impact of key factors on identification accuracy.}
\label{fig:factors}
\end{figure}

As explained in \S\ref{sec:attr}, our dataset contains labeled metadata for a number of user attributes, including device information and some basic demographics. While we avoided using this data in our identification model in order to achieve purely motion-based identification, we later used all of this information to perform a key factor analysis so as to better understand which attributes affect the identifiability of a user. The 15 most important factors are summarized in Fig. \ref{fig:factors}. This summary evaluates the impact of each factor on the accuracy of layers 1 and 2, as not all users are present in layer 3.

Fig. \ref{fig:factors} reveals some interesting trends with respect to the factors which most impacted identification accuracy.
Some devices, such as Windows Mixed Reality, are less conducive to identification, perhaps due to an overreliance on low-quality dead reckoning for tracking. Others, like Valve Index, yield better than average user identification, which may be due to their highly precise outside-in tracking system.

Users from certain countries, particularly Japan and South Korea, are significantly easier to identify, implying there may be detectable cultural differences in play style. This result is highly statistically significant, with over 99\% identification accuracy for users from those two countries.

\begin{figure}[H]
\includegraphics[width=\linewidth]{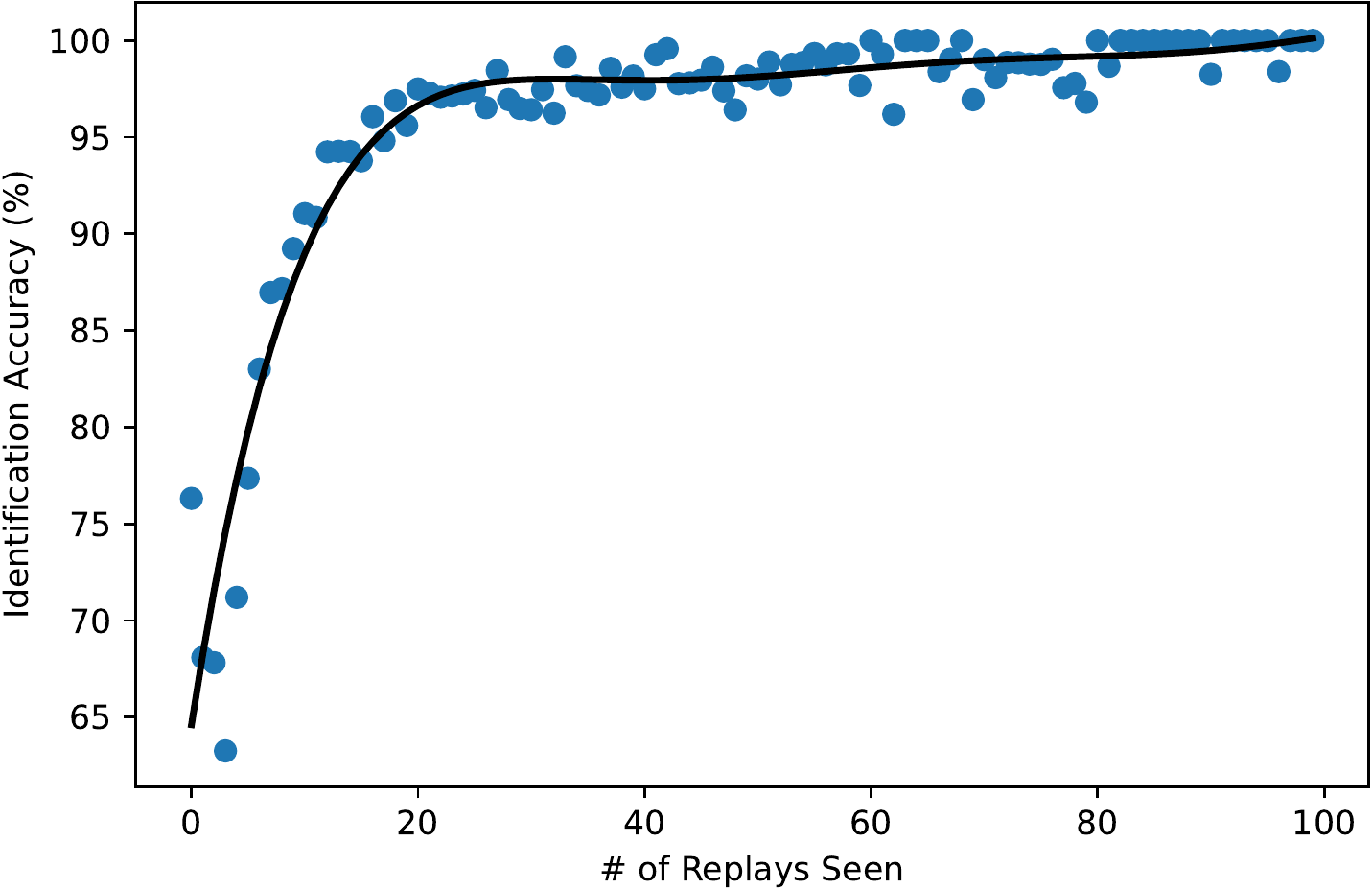}
\centering
\caption{Replays per user vs. identification accuracy.}
\label{fig:replays}
\end{figure}

However, by far the most important factor in determining identification accuracy is the number of total replays observed from a target user, regardless of how many samples were actually used to train the model. Users with 5 or less total replays submitted were significantly harder to identify, while the 5,000 or so users with 100 or more replays could be identified with over 99.5\% accuracy. The identification accuracy for users is charted against the number of replays in Fig. \ref{fig:replays}.

The clear trend of users with more replays (and thus more time spent in the game) being more easily identifiable is indicative of something other than more data being available, as the full 150 training features can easily be extracted from a single 5-minute session. Rather, it suggests that users with more experience are likely to develop a distinct play style (and reinforce the corresponding muscle memory) over time.
Highly experienced players are thus more likely than novices to exhibit a repeatable response to the same stimulus, with veteran users becoming so consistent in their movements that they can be identified with near-perfect accuracy.
\section{Explanations}
\label{sec:explanations}

\begin{figure}[H]
\includegraphics[width=\linewidth]{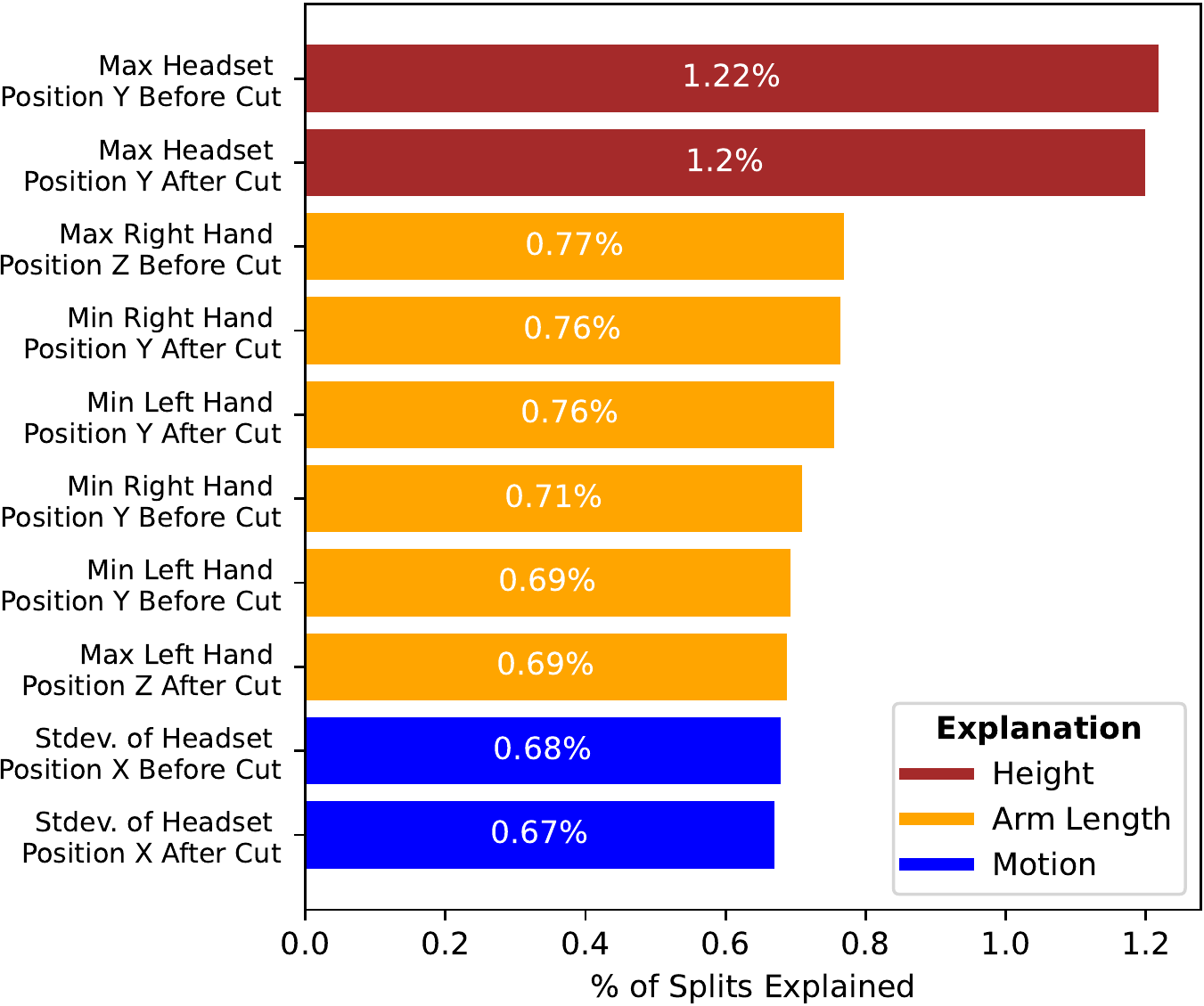}
\centering
\caption{Explanation for 10 most important features.}
\label{fig:importance}
\end{figure}

An additional benefit of using a LightGBM model rather a deep learning method is the relative ease of explaining the importance of each feature. Fig. \ref{fig:importance} shows the percentage of splits attributable to each of the 10 most important features (out of 232) in our final identification model.

As illustrated in Fig. \ref{fig:importance}, many of the most important features for identification correspond to obvious physical measurements. For example, the two most important features, which measure the maximum Y-position of the headset before and after the cut, are an obvious proxy for the user's height (and posture). Similarly, the next six most important features seemingly measure the length of the user's arms when furthest outstretched. These first eight features alone account for 6.8\% of the splits and 10.2\% of the gain of the identification model, providing about 12 bits of real entropy -- enough information to accurately identify as many as 4,000 users.


It is no coincidence that these easily understandable features are by far the most important for identification.
Unlike motion features, which are highly dependent on the specific action being taken, features that measure some static physical dimension of a user are highly consistent throughout a replay and across sessions. Thus, while the importance of any given motion feature may vary depending on the context of a sample, models can be sure to glean some information from the static features of every sample, regardless of context.

\begin{figure}[H]
\includegraphics[width=\linewidth]{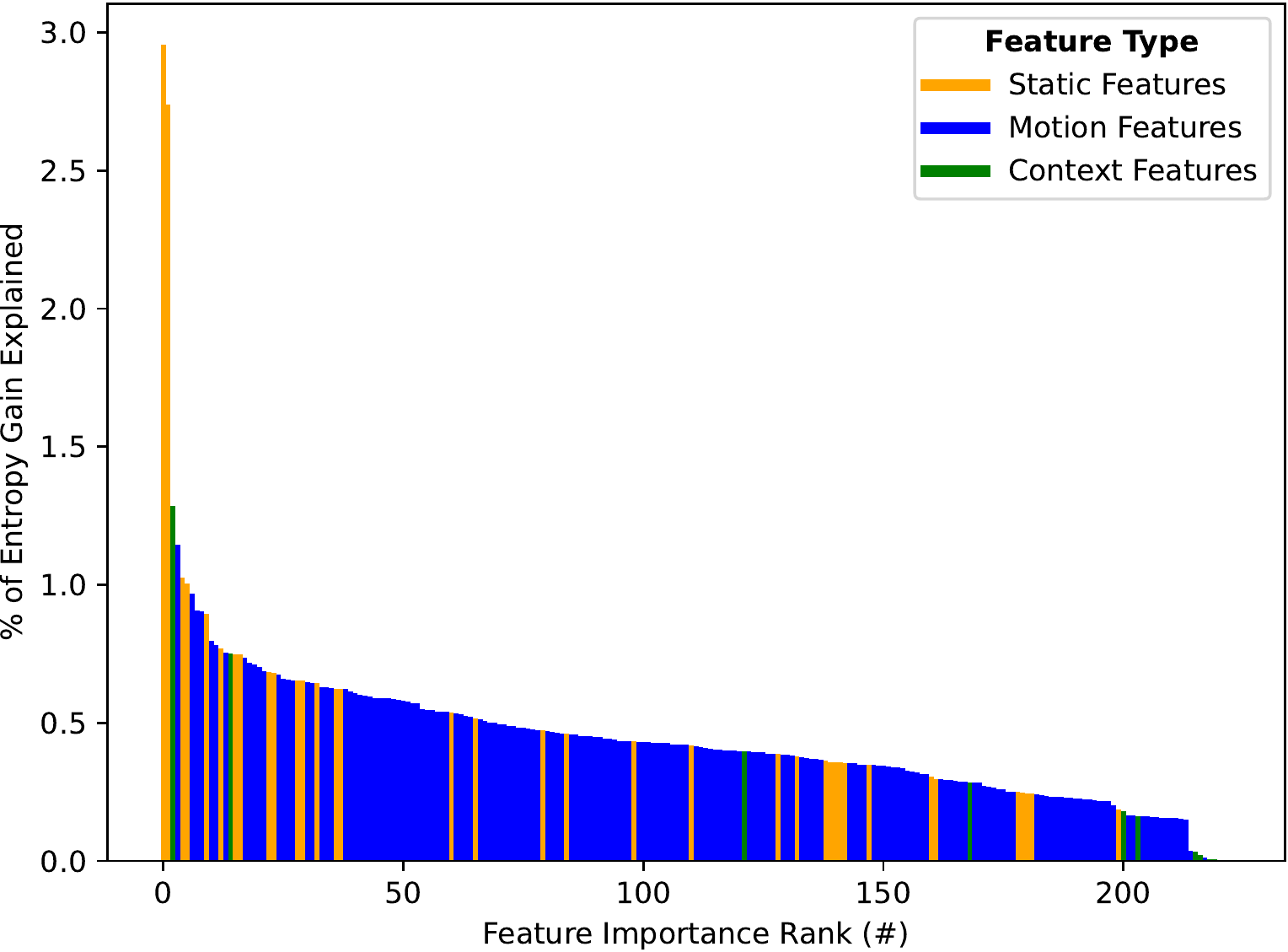}
\centering
\caption{Entropy explained by all feature types.}
\label{fig:features}
\end{figure}

Still, these simple measurements alone hardly account for the identification of 50,000 users. A more complete picture is provided by Fig. \ref{fig:features}, which shows the percentage of overall information gain explained by all 232 utilized features.

As is evident in Fig. \ref{fig:features}, motion features actually play a major role in identifying users. While static measurements comprise many of the most important features, they account for only 22.9\% of the overall performance of the model. Motion features constitute 73.9\% of all entropy gain, while contextual features compose the remaining 3.2\%.
Clearly, motion features actually represent the majority of information used by our identification model, and the task of identifying over 50,000 users would not have been possible without them.
\section{Discussion}
\label{sec:discussion}

In light of the fundamental factors working against this result, the identification accuracy achieved by this paper may even be stronger than it initially appears. Unlike the laboratory studies with which this work can be most directly compared, our study endures many of the pitfalls associated with utilizing a dataset from ``in the wild.'' Chief among them is the fact that many users may actually have more than one account or play on multiple devices, resulting in the presence of multiple distinct classes which are in fact identical.

Furthermore, our definition of a ``session'' is more rigorous than the related works, with training and testing data for users originating from completely separate days in almost all cases. The largest comparable study (Miller et al. \cite{miller_personal_2020}) records 10 short sessions of a user on the same day. Therefore, our results represent the consistent identification of a user across wider periods of time, a task that is far more difficult than correlating motion segments recorded in rapid succession.

Lastly, we argue that our work was the first to fully and demonstrably leverage actual movement for identification in VR. As demonstrated in \S\ref{sec:explanations}, deriving simple measurements like height and arm lengths are sufficient for a model to identify tens or even hundreds of users, as is seen in Miller et al. \cite{miller_personal_2020}. This speculation is supported by the fact that users in that study were instructed to simply observe a number of 360-degree VR videos, a relatively static task that does not fundamentally implicate much movement.
By contrast, identifying 50,000 users would not have been possible without leveraging actual motion patterns, which was made possible by our featurization approach which contextualizes observed motion relative to relevant virtual objects involved in a repeatable activity. The model explainability results of \S\ref{sec:explanations} indicate that motion features played a key role in identifying users, accounting for a majority of the model's information gain.

\subsection{Limitations}
There are a few notable limitations to the work presented in this paper.
Most importantly, several features were used to identify users that are arguably unique to the Beat Saber application. While Beat Saber is currently the most popular VR application in existence, it is not clear, without further investigation, whether these results will generalize to other types of VR applications.
Furthermore, the ``ground truth'' values for some of the attributes reported in \S\ref{app:distribution}, namely height and handedness, are based on user-configurable settings, and as such, should be treated as self-reported. Indeed, many players are known to deliberately misconfigure their height setting to obtain a perceived performance advantage. 

As described in \S\ref{sec:related} and quantified in \S\ref{sec:alg}, deep learning models, though broadly desirable, empirically underperformed tree-based models in our experiments.
We found the identification performance of traditional ML models to be sufficient in light of the main focus of this paper, which is to shed light on the sheer magnitude of the privacy concerns implicated by collecting telemetry data in VR applications.  An advantage of using LightGBM in this setting is the ability to generate rich model explanations, as presented in \S\ref{sec:explanations}.

\vspace{-0.5em}
\subsection{Future Work}
For the reasons discussed above, our results rely on tree-based models rather than using deep learning. In the future, we hope to see deep learning models (especially advanced sequential models like transformer-based models~\cite{devlin2018bert}) applied to the same problem, perhaps enabled by a combination of distributed machine learning and more efficient techniques.

There are several interesting applications of our results to Beat Saber specifically, as well as VR gaming in general. These include advanced cheating detection, score prediction, skill-based matchmaking, and map recommendation engines.

By collecting surveys to measure ground truth, future work could aim to infer specific attributes from VR telemetry, including demographics, biometrics, and perhaps even medical conditions, turning VR into a useful measurement tool.

Finally, and perhaps most importantly, we hope to motivate future work into defensive applications and techniques. We hope to see future works which intelligently corrupt VR replays to obscure identifiable properties without impeding their original purpose (e.g., scoring or cheating detection).
\vspace{-0.5em}
\section{Conclusion}
\label{sec:conclusion}

While perhaps not surprising to experts in biomechanics, the extent to which users can be uniquely identified by observing just a few seconds of motion of their head and hands may indeed be surprising to most. Though we don't presently think of movement patterns as a uniquely identifiable characteristic to the same extent as faces and fingerprints, results like those presented in this paper may serve to change this assumption.
Researchers have long speculated that individuals might be identifiable by their movements on a much larger scale than lab studies are able to demonstrate, but datasets with motion from tens of thousands of users did not begin to emerge until the recent widespread adoption of VR.

As we slowly realize the increasing role that virtual reality and the ``metaverse'' may soon play in our lives, more attention should be given to the security and privacy implications of these platforms. The same telemetry streams which are essential to their operation should in fact be considered highly sensitive data that may reveal a plethora of information about an end user.
We hope to motivate further research into privacy-preserving technologies which may be deployed to enable the use of VR without revealing private user information.

\section*{Acknowledgments}
We would like to acknowledge and thank Xiaoyuan Liu, Charles Dove, Julien Piet, Mark Roman Miller, Gonzalo Munilla Garrido, Ines Bouissou, Beni Issler, Eric Wallace, and Yu Gai for their advice, support, and assistance. We additionally thank Beat Games, ScoreSaber, BeatLeader, and their respective teams, for providing access to the data used in this study. We particularly thank Viktor Radulov and Dziugas Ramonas for their ongoing guidance. This work was supported in part by the National Science Foundation, by the National Physical Science Consortium, by the Fannie and John Hertz Foundation, and by the Berkeley Center for Responsible, Decentralized Intelligence. Any opinions, findings, and conclusions or recommendations expressed in this material are those of the authors and do not necessarily reflect the views of their employer or the supporting entities. We sincerely thank all of the VR users whose published data made this work possible.

\section*{Availability}

The featurization, normalization, training, and testing scripts used in this paper are available for review in our public GitHub repository, along with detailed logs, outputs, and results:

\vspace{1em}

\centerline{
    \url{https://github.com/MetaGuard/Identification}
}

\vspace{1em}

\noindent For privacy and logistical reasons, we have not published the full training dataset. Researchers may contact the authors if they wish to obtain a copy for replication purposes.


\bibliographystyle{plainurl}
\bibliography{900-References}

\appendix
\section{LightGBM Hyperparameters}
\label{app:params}

\begin{itemize}[leftmargin=*,itemsep=0em]
    \item objective=`multiclass'
    \item boosting\_type=`goss'
    \item colsample\_bytree=0.6933333333333332
    \item learning\_rate=0.1
    \item max\_bin=63
    \item max\_depth=-1
    \item min\_child\_weight=7
    \item min\_data\_in\_leaf=20
    \item min\_split\_gain=0.9473684210526315
    \item n\_estimators=200
    \item num\_leaves=33
    \item reg\_alpha=0.7894736842105263
    \item reg\_lambda=0.894736842105263
\end{itemize}
\clearpage

\section{Participant Distribution}
\label{app:distribution}

\noindent \textbf{Replays} \dotfill \textbf{~55,541} \\
$\leq 5$ \dotfill ~14,945 (26.9\%) \\
$6$--$10$ \dotfill ~8,639 (15.6\%) \\
$11$--$24$ \dotfill ~12,495 (22.5\%) \\
$25$--$99$ \dotfill ~14,012 (25.2\%) \\
$\geq  100$ \dotfill ~5,450 (9.8\%) \\

\noindent \textbf{Platform} \dotfill \textbf{~55,541} \\
SteamVR \dotfill ~42,035 (75.7\%) \\
Oculus \dotfill ~11,269 (20.3\%) \\
Oculus PC \dotfill ~2,223 (4.0\%) \\
Others \dotfill ~14 (0.0\%) \\

\noindent \textbf{Runtime} \dotfill \textbf{~55,541} \\
OpenVR \dotfill ~42,039 (75.7\%) \\
Oculus \dotfill ~13,492 (24.3\%) \\
Unknown \dotfill ~10 (0.0\%) \\

\noindent \textbf{Headset} \dotfill \textbf{~55,541} \\
Oculus Quest 2 (Standalone) \dotfill ~25,857 (46.6\%) \\
Oculus Quest 2 (Quest Link) \dotfill ~4,124 (7.4\%) \\
Valve Index \dotfill ~8,820 (15.9\%) \\
Oculus Rift S \dotfill ~4,483 (8.1\%) \\
HTC Vive \dotfill ~2,408 (4.3\%) \\
Oculus Rift CV1 \dotfill ~2,061 (3.7\%) \\
Pico Neo 3 \dotfill ~1,595 (2.9\%) \\
Oculus Quest (Standalone) \dotfill ~1,453 (2.6\%) \\
Oculus Quest (Quest Link) \dotfill ~313 (0.6\%) \\
PICO 4 \dotfill ~905 (1.6\%) \\
HTC VIVE Pro \dotfill ~728 (1.3\%) \\
HP Reverb G20 \dotfill ~644 (1.2\%) \\
HTC Vive Cosmos Elite \dotfill ~395 (0.7\%) \\
HTC VIVE Pro 2 \dotfill ~328 (0.6\%) \\
Samsung Windows Mixed Reality \dotfill ~304 (0.5\%) \\
HTC Vive Cosmos \dotfill ~226 (0.4\%) \\
Others \dotfill ~897 (1.6\%) \\

\noindent \textbf{Controller} \dotfill \textbf{~55,541} \\
Oculus Quest Controller \dotfill ~16,449 (29.6\%) \\
Oculus Touch Controller \dotfill ~11,240 (20.2\%) \\
Valve Knuckles Controller \dotfill ~9,805 (17.7\%) \\
Oculus Rift S Controller \dotfill ~3,202 (5.8\%) \\
HTC Vive Controller \dotfill ~1,958 (3.5\%) \\
Pico Neo 3 Controller \dotfill ~1,443 (2.6\%) \\
Oculus Rift CV1 Controller \dotfill ~1,265 (2.3\%) \\
Oculus Quest Controller \dotfill ~665 (1.2\%) \\
HTC VIVE Pro Controller \dotfill ~602 (1.1\%) \\
Others \dotfill ~8,912 (16.0\%) \\

\newpage

\noindent \textbf{Handedness} \dotfill \textbf{~55,541} \\
Right \dotfill ~53,144 (95.7\%) \\
Left \dotfill ~2,397 (4.3\%) \\

\noindent \textbf{Height} \dotfill \textbf{~55,541} \\
$\leq 1.5$~m \dotfill ~4,888 (8.8\%) \\
$1.5$~m -- $1.6$~m \dotfill ~4,721 (8.5\%) \\
$1.6$~m -- $1.7$~m \dotfill ~17,273 (31.1\%) \\
$1.7$~m -- $1.8$~m \dotfill ~18,495 (33.3\%) \\
$1.8$~m -- $1.9$~m \dotfill ~6,720 (12.1\%) \\
$\geq 1.9$~m \dotfill ~3,444 (6.2\%) \\

\noindent \textbf{Countries} \dotfill \textbf{~55,541} \\
US \dotfill ~15,142 (27.3\%) \\
DE \dotfill ~2,404 (4.3\%) \\
GB \dotfill ~2,350 (4.2\%) \\
CN \dotfill ~1,964 (3.5\%) \\
CA \dotfill ~1,563 (2.8\%) \\
JP \dotfill ~1,337 (2.4\%) \\
AU \dotfill ~988 (1.8\%) \\
FR \dotfill ~955 (1.7\%) \\
NL \dotfill ~767 (1.4\%) \\
RU \dotfill ~743 (1.3\%) \\
PL \dotfill ~650 (1.2\%) \\
HK \dotfill ~545 (1.0\%) \\
BR \dotfill ~349 (0.6\%) \\
CZ \dotfill ~344 (0.6\%) \\
FI \dotfill ~335 (0.6\%) \\
KR \dotfill ~304 (0.5\%) \\
NO \dotfill ~297 (0.5\%) \\
SE \dotfill ~288 (0.5\%) \\
ES \dotfill ~282 (0.5\%) \\
AT \dotfill ~277 (0.5\%) \\
DK \dotfill ~255 (0.5\%) \\
SG \dotfill ~241 (0.4\%) \\
BE \dotfill ~201 (0.4\%) \\
IT \dotfill ~188 (0.3\%) \\
NZ \dotfill ~159 (0.3\%) \\
TW \dotfill ~157 (0.3\%) \\
MX \dotfill ~137 (0.2\%) \\
CH \dotfill ~116 (0.2\%) \\
HU \dotfill ~114 (0.2\%) \\
CL \dotfill ~111 (0.2\%) \\
IL \dotfill ~101 (0.2\%) \\
TH \dotfill ~88 (0.2\%) \\
AR \dotfill ~88 (0.2\%) \\
IE \dotfill ~86 (0.2\%) \\
UA \dotfill ~85 (0.2\%) \\
PT \dotfill ~76 (0.1\%) \\
Others \dotfill ~21389 (38.5\%) \\

\end{document}